\documentclass[12pt]{iopart}

\usepackage{iopams} 
\usepackage{graphicx}

\usepackage{hyperref}
\hypersetup{colorlinks=false} 

\begin{document}

\title[Radon induced background processes in the KATRIN pre-spectrometer]{Radon induced background processes in the KATRIN pre-spectrometer}

\author{F. M. Fr\"ankle$^1$ $^2$ $^3$, L. Bornschein$^1$, G. Drexlin$^1$, F. Gl\"uck$^1$ $^4$, S. G\"orhardt$^1$, W. K\"afer$^1$, S. Mertens$^1$, N. Wandkowsky$^1$, J. Wolf$^1$}

\address{1 Karlsruhe Institute of Technology, 76131 Karlsruhe, Germany}
\address{2 Department of Physics, University of North Carolina, Chapel Hill, NC, USA}
\address{3 Triangle Universities Nuclear Laboratory, Durham, NC, USA}
\address{4 KFKI, RMKI, H-1525 Budapest, POB 49, Hungary}
\ead{florian.fraenkle@kit.edu}

\begin{abstract}
The \textbf{KA}rlsruhe \textbf{TRI}tium \textbf{N}eutrino (KATRIN) experiment is a next generation, model independent, large scale tritium $\beta$-decay experiment to determine the effective electron anti-neutrino mass by investigating the kinematics of tritium $\beta$-decay with a sensitivity of 200~meV/c$^2$ using the MAC-E filter technique. In order to reach this sensitivity, a low background level of 10$^{-2}$~counts per second (cps) is required. This paper describes how the decay of radon in a MAC-E filter generates background events, based on measurements performed at the KATRIN pre-spectrometer test setup. Radon (Rn) atoms, which emanate from materials inside the vacuum region of the KATRIN spectrometers, are able to penetrate deep into the magnetic flux tube so that the $\alpha$-decay of Rn contributes to the background. Of particular importance are electrons emitted in processes accompanying the Rn $\alpha$-decay, such as shake-off, internal conversion of excited levels in the Rn daughter atoms and Auger electrons. While low-energy electrons ($<$ 100~eV) directly contribute to the background in the signal region, higher energy electrons can be stored magnetically inside the volume of the spectrometer. Depending on their initial energy, they are able to create thousands of secondary electrons via subsequent ionization processes with residual gas molecules and, since the detector is not able to distinguish these secondary electrons from the signal electrons, an increased background rate over an extended period of time is generated.
\end{abstract}

\maketitle

\section{Introduction}
Neutrinos play an important role in the evolution of the universe. Due to their large abundance - the neutrino density of the universe is about 330~$\nu$'s/cm$^3$ for all three flavours - neutrinos can affect the evolution of large scale structures depending on their mass. Neutrino oscillation experiments, have shown that neutrinos are massive, and set a lower limit of 40~meV/c$^2$ \cite{pdg} on the largest neutrino mass. Experiments investigating the kinematics of $\beta$-decay set an upper limit on the effective electron neutrino mass of 2.3~eV/c$^2$ \cite{kra05} \cite{lob99}. The \textbf{KA}rlsruhe \textbf{TRI}tium \textbf{N}eutrino (KATRIN) experiment \cite{kat04} uses a model independent method to determine the mass of the electron anti-neutrino ($m_{\overline{\nu}_e}$) by investigating the kinematics of tritium $\beta$-decay. With a sensitivity of 200~meV/c$^2$ on $m_{\overline{\nu}_e}$, KATRIN has the potential to probe the mass range of $m_{\overline{\nu}_e}$ where neutrinos have a significant influence on structure formation in the universe. In order to achieve this sensitivity, a background rate of less than 10$^{-2}$ counts per second (cps) close to the $\beta$-decay endpoint at 18.6~keV is required. 

\subsection{KATRIN experimental setup}

\begin{figure}[h]
 \centering
 \includegraphics[width=150mm]{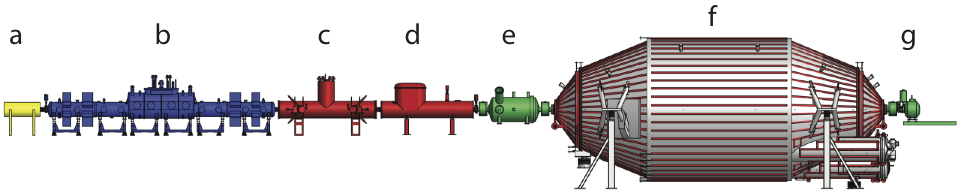}
 \caption{Schematic overview of the KATRIN experimental setup: \textbf{a} rear section, \textbf{b} windowless gaseous molecular tritium source (WGTS), \textbf{c} differential pumping section (DPS), \textbf{d} cryogenic pumping section (CPS), \textbf{e} pre-spectrometer, \textbf{f} main spectrometer, \textbf{g} detector system. The overall setup has a length of about 70~m.}
 \label{fig:katrin}
\end{figure} 

The KATRIN setup (see figure \ref{fig:katrin}) consists of a high luminosity windowless gaseous molecular tritium source (WGTS), a differential and cryogenic pumping system for tritium retention with super-conducting magnets for electron transport, and a tandem spectrometer section (pre-spectrometer and main spectrometer) for energy analysis, followed by a detector system for counting transmitted $\beta$-electrons. KATRIN is designed so that electrons in a magnetic flux tube of 191~Tcm$^2$ are guided adiabatically from the WGTS to the detector system. In order to achieve the desired sensitivity, the WGTS - in which tritium decays with an activity of about 10$^{11}$~Bq - needs to be stable on the 0.1~\% level with respect to injection pressure and temperature, with an absolute temperature of about 30~K. The main spectrometer (length 24~m, diameter 10~m), which works as a retarding electrostatic spectrometer of MAC-E\footnote{\textbf{M}agnetic \textbf{A}diabatic \textbf{C}ollimation combined with an \textbf{E}lectrostatic Filter.} filter type \cite{bea80}, will have an energy resolution of 0.93~eV at the tritium endpoint (18.6~keV). For the high precision energy analysis, a 1~ppm stability is needed for the \mbox{-18.6}~kV retarding potential.

\subsection{MAC-E filter}

\begin{figure}[h]
 \centering
 \includegraphics[width=150mm]{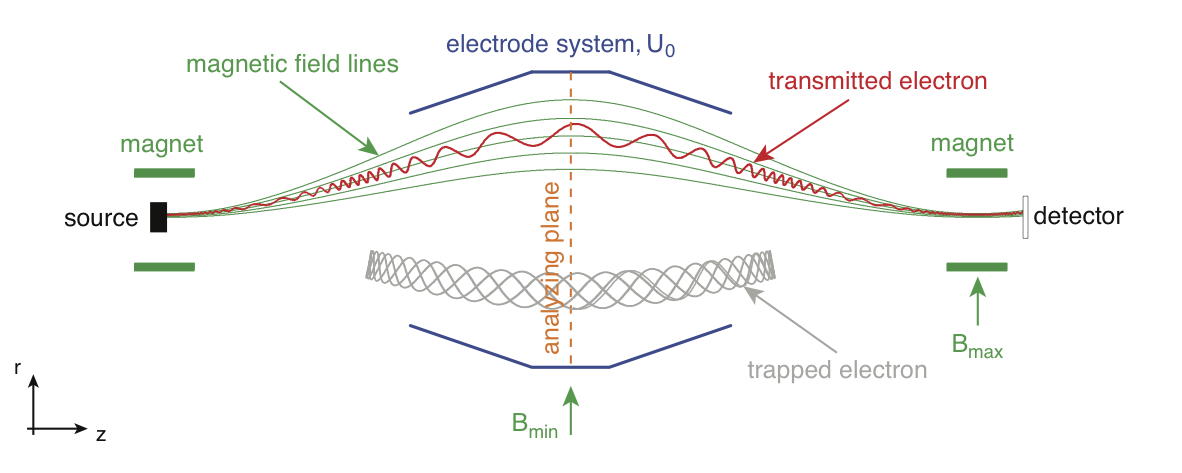}
 \caption{Schematic drawing of a MAC-E filter with guiding magnetic field lines, source, electrostatic spectrometer and detector. The maximum of the retarding potential $U_0$ defines the analyzing plane. The trajectories of a transmitted electron and a magnetically trapped electron are shown.}
 \label{fig:psmace}
\end{figure} 

In order to perform a high precision energy analysis of the $\beta$-decay electrons in the KATRIN experiment, the so-called MAC-E filter technique is used, which is able to combine high luminosity with high energy resolution. In general, a MAC-E filter consists of two solenoids that create a strong magnetic guiding field for electrons\footnote{In principle, a MAC-E filter is not limited to electrons and works with any electrically charged particle, as long as the magnetic guidance is adiabatic.} (see figure \ref{fig:psmace}). This magnetic field $B$ has a minimum $B_{min}$ between the coils and a maximum $B_{max}$ inside the centre of one coil. For adiabatic motion the orbital magnetic moment $\mu = E_{\perp}/B$ of electrons remains constant, so that their transversal energy component $E_{\perp}$ is transformed into longitudinal energy as they move towards $B_{min}$. The actual energy analysis is done with an electrostatic retarding potential at $B_{min}$, which is created by an electrode system operated on a variable scanning potential $U_0$. Thus, a MAC-E filter acts as an integrating high pass filter with a finite energy resolution $\Delta E = U_0\cdot B_{min}/B_{max}$.

The magnetic field of a MAC-E filter provides an intrinsic protection against background created by electrically charged particles originating from the electrodes. However, if low energy electrons are produced inside the volume of a MAC-E filter via ionization of residual gas molecules or radioactive decays, these electrons are accelerated by the retarding potential to the energy of the actual signal electrons. Therefore, processes that create such electrons need to be prevented in order to achieve a low background level.

\section{Pre-spectrometer test setup}

The pre-spectrometer test setup \cite{hab09} \cite{fra10} acts as a prototype for the larger main spectrometer in order to test the vacuum concept and the new electromagnetic design of the KATRIN spectrometers. In addition, new technologies are tested and developed at the test setup, which will later be applied to KATRIN (e.g. an active HV stabilization, data acquisition system and background suppression techniques).
The test setup consists of the pre-spectrometer vessel, an ultra high vacuum recipient including an inner electrode system, two super-conducting solenoids (max. 4.5~T), a photoelectric electron source (e-gun) and a detector chamber, housing a monolithic 64 pixel silicon PIN diode (SPD).

\subsection{Pre-spectrometer}

\begin{figure}[h]
 \centering
 \includegraphics[width=150mm]{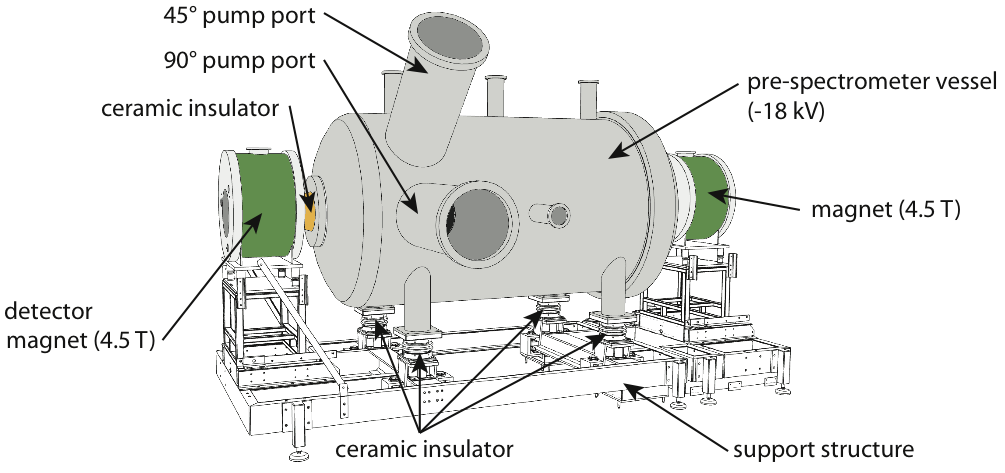}
 \caption{The pre-spectrometer vessel is mounted on a stainless steel support structure and is electrically isolated from its surroundings via ceramic insulators. On both sides, a superconducting magnet coil is attached, providing an axial field at the centre of the coil of $B$~=~4.5~T.}
 \label{fig:ps3d}
\end{figure} 

Figure \ref{fig:ps3d} shows a schematic overview of the pre-spectrometer vessel and the super-conducting solenoids. The pre-spectrometer vessel is made of 10~mm thick stainless steel of type 316LN. It has a length of 3.38~m and an inner diameter of 1.68~m. Two pump ports with a length of 1~m and a diameter of 0.5~m are welded to the vessel at 45$^{\circ}$ and 90$^{\circ}$ with respect to the vertical $y$-axis. The total volume of the pre-spectrometer vacuum chamber is 8.5~m$^3$.

\begin{figure}[h]
 \centering
 \includegraphics[width=150mm]{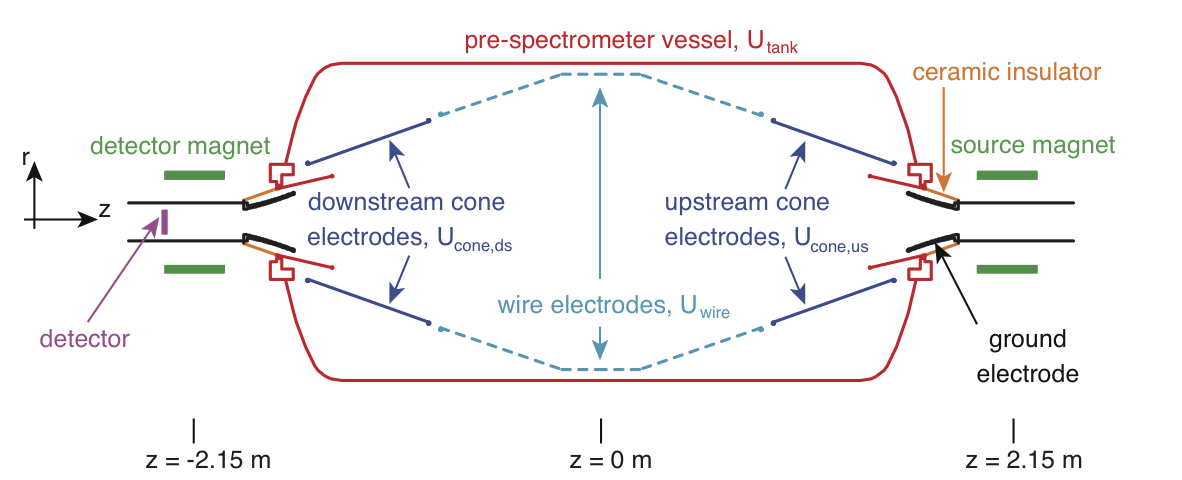}
 \caption{Schematic overview of pre-spectrometer electrode system.}
 \label{fig:pselectrodes}
\end{figure}

The pre-spectrometer vessel is electrically isolated from its surroundings and can be set to a maximum potential of -35~kV. Inside the pre-spectrometer vessel there are three sets of conical axially symmetric inner electrodes (upstream cone, wire, downstream cone, see figure \ref{fig:pselectrodes}). The cone electrodes are made of full metal sheets and the central wire electrode is made of 240 wires of 0.5~mm diameter, mounted in the axial direction. At both ends, conical electrodes held at ground potential provide the counterpart for the retarding field.

The mechanical vacuum system of the pre-spectrometer is attached to the 90$^{\circ}$ pump port. It consists of a cascaded set of turbo-molecular pumps (TMPs) with an effective pumping speed of 1000~l/s and a dry fore-pump. The main pumping speed for hydrogen is produced by a non-evaporable getter (NEG) pump, installed inside the 45$^{\circ}$ pump port, consisting of 90~m of SAES St707 getter strips with an effective pumping speed of 25000~l/s. A pressure in the 10$^{-11}$~mbar range can be reached at room temperature \cite{bor06} after bake-out of the system at 200$^{\circ}$C and activation of the getter at 350$^{\circ}$C.

\subsection{Segmented pixel detector}
In order to detect electrons originating from the pre-spectrometer or the electron source, a large quadratic monolithic silicon PIN diode \cite{wue06} (sensitive area 16~cm$^2$), which is segmented into 8~$\times$~8 individual quadratic pixels of equal area, is attached on one side of the pre-spectrometer. To be able to cover any position within the flux tube, the detector is mounted on a manipulator that is movable in all three directions. At the default measurement position, the detector is centered in the flux tube at a magnetic field of 3.4~T and covers 28.5~\% of the flux tube. A customized system is used for data acquisition \cite{wue10}.

\section{Pre-spectrometer background measurement results}
In order to characterize the background behavior of the pre-spectrometer, several measurements at the design values of the magnetic field (4.5~T) and electric potential of the outer vessel \mbox{(-18~kV)} were performed \cite{fra10}. The inner electrodes were kept on -18.5~kV in order to take advantage of the electrostatic shielding of the wire electrode against low energetic background electrons originating from the walls of the pre-spectrometer vessel \cite{fla04}.

\begin{figure}[h]
 \centering
 \includegraphics[width=140mm]{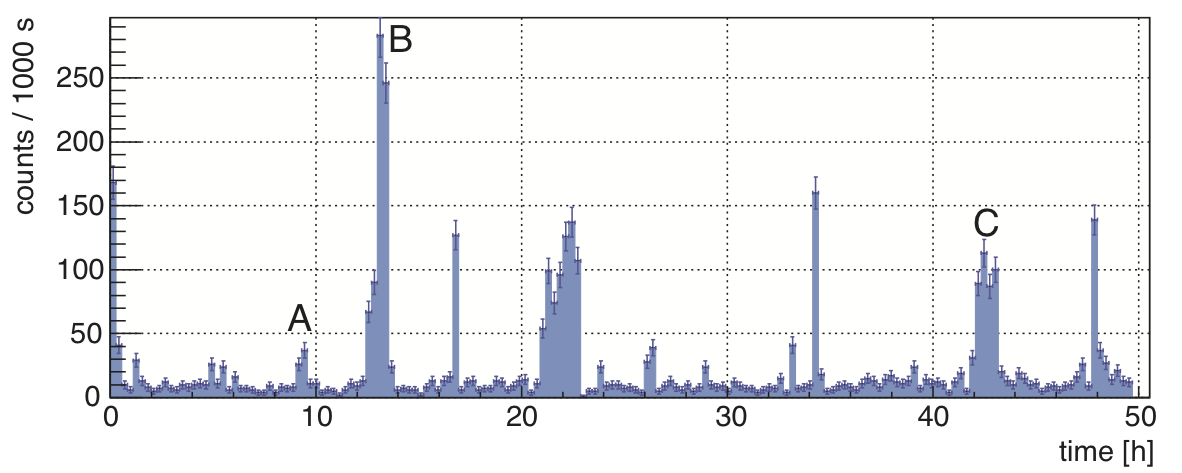}
 \caption{Rate versus time plot for a 50~h background measurement with the pre-spectrometer standard operating conditions for a magnetic field of 4.5~T and an electric potential of -18.5~kV at a pressure of 10$^{-10}$~mbar. The rates include the intrinsic detector background of (6.3~$\pm$~0.2)~$\cdot$~10$^{-3}$~cps. The width of a time bin is 1000~s. The pixel distributions for the intervals with increased rates A to C are shown in figure \ref{fig:rings}.}
 \label{fig:rate}
\end{figure}

\begin{figure}[p]
 \centering
 \includegraphics[width=140mm]{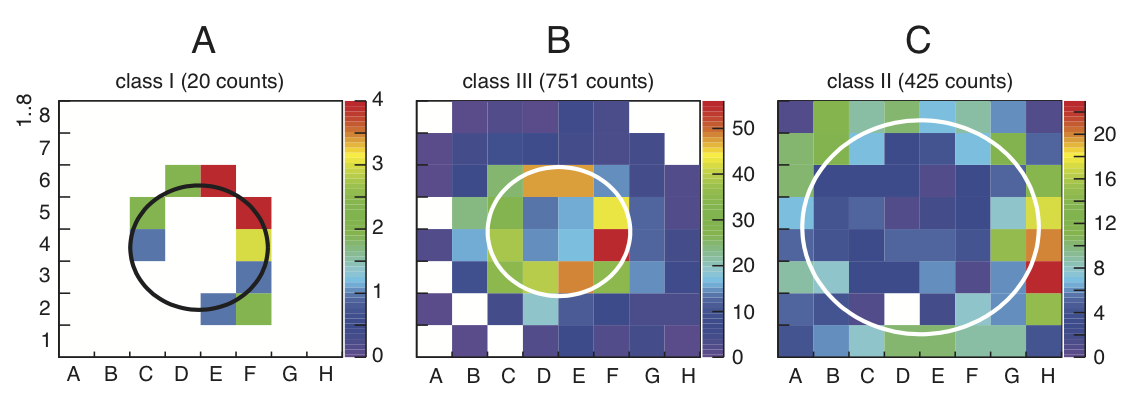}
 \caption{Pixel distributions for different time intervals with increased rate (see figure \ref{fig:rate}), together with the ring radius resulting from a fit. The ring radius varies along the spectrometer as $r_1/r_2 = \sqrt{B_2/B_1}$ and is calculated to be 24~cm in the analyzing plane for ring C.}
 \label{fig:rings}
\end{figure}

\begin{figure}[p]
 \centering
 \includegraphics[width=140mm]{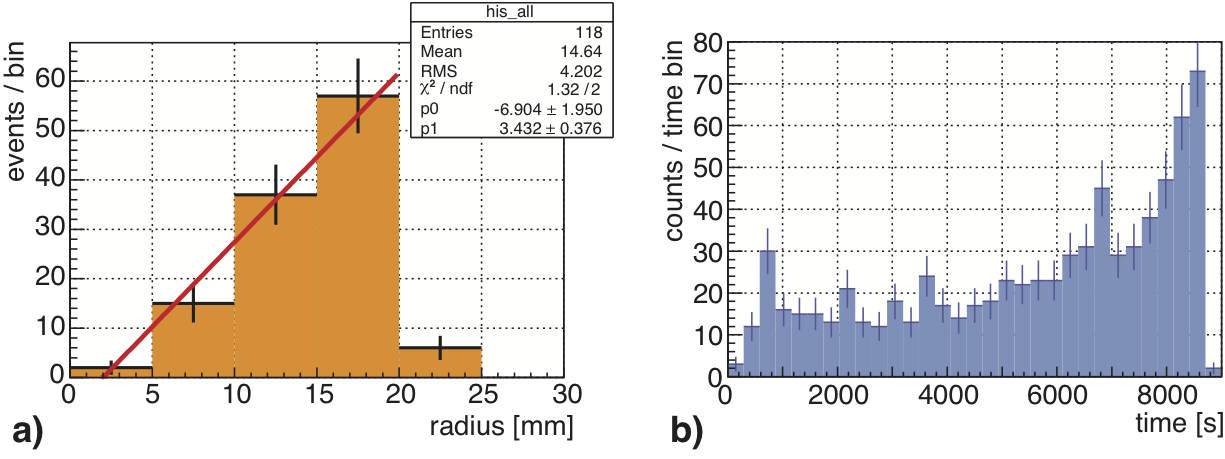}
 \caption{\textbf{a)} Distribution of ring radii, the bin size corresponds to the width of a detector pixel (5~mm). \textbf{b)} Time structure of an interval of elevated rate. Each time bin has a width of 300~s. This event lasted more than two hours and, towards the end of the event, a significant increase in rate was observed.}
 \label{fig:distribution}
\end{figure} 

Figure \ref{fig:rate} shows the count rate versus time at the detector in an energy interval between 15 and 21~keV for a 50~h background measurement at a pressure of 10$^{-10}$~mbar. Most of the time the rate is close to the level of the intrinsic detector background of (6.3~$\pm$~0.2)~$\cdot$~10$^{-3}$~cps. However, elevated rates (up to 250~$\cdot$~10$^{-3}$~cps) are observed, which last between $<$15 and 120~min. 

The pixel distributions of three selected intervals of increased rate are shown in figure \ref{fig:rings} (A to C). These background counts are not distributed randomly on the detector, but form distinct ring patterns. The concentric rings have different radii, and the number of single detector counts belonging to one ring varies between 10 and about 2200 counts. The energy spectra of the ring events show peaks around 18.5~keV that correspond to the potential of the inner electrode system. From this, it can be concluded that low energetic electrons are produced in the volume of the pre-spectrometer, which are accelerated to 18.5~keV by the electrode potential as they follow the magnetic field lines towards the detector.

The ring events are classified into four different event classes (see table \ref{tab:eventclass}), depending on the number of single counts at the detector during the time of elevated rate. Only class \emph{I} to \emph{III} events can be identified reliably as ring structures, while class \emph{0} events contribute as a diffuse background component. The distribution of ring radii is shown in figure \ref{fig:distribution}a). The linear increase towards larger radii is consistent with the assumption that the starting points of ring events are distributed homogeneously inside the pre-spectrometer, because the volume element $dV(r)$ increases with $r dr$.

\begin{table}
	\caption{Definition of ring structure event classes, detector counts: number of detector counts during an interval of increased rate (see figure \ref{fig:rate}), primary energy: order of magnitude of the initial energy of a magnetically trapped electron.}
	\label{tab:eventclass}
		\begin{indented}
		\item[]
		\begin{tabular}{@{}lll}
			\br
			event class & detector counts & primary energy\\
			\mr
			\emph{0} & 1 .. 9 & 0.1~keV\\
			\emph{I} & 10 .. 50 & 1~keV\\
			\emph{II} & 51 .. 500 & 10~keV\\
			\emph{III} & 501 .. 5000 & 100~keV\\
			\br
		\end{tabular}
	\end{indented}
\end{table}

\begin{figure}[h]
 \centering
 \includegraphics[width=140mm]{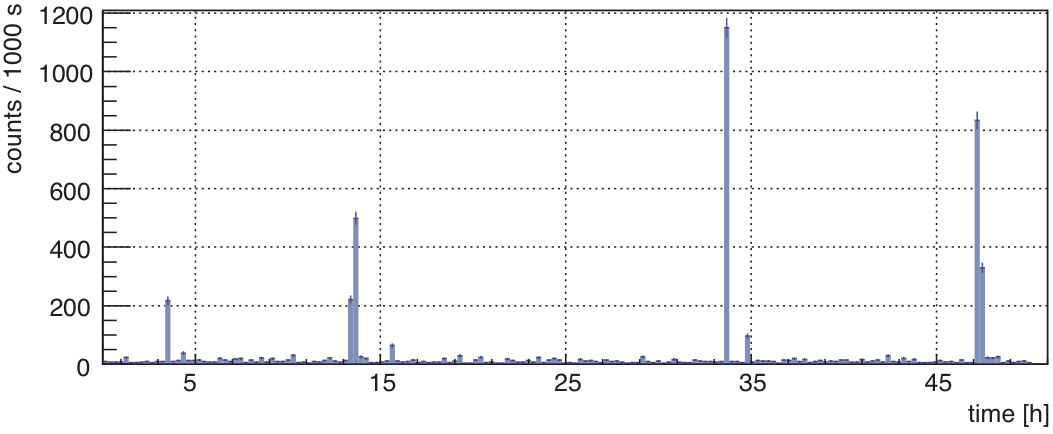}
 \caption{Rate versus time plot for a 50~h background measurement with the same experimental settings as in figure \ref{fig:rate}, but with an increased pressure of $2\cdot10^{-9}$~mbar.}
 \label{fig:rate5to55h}
\end{figure}

Figure \ref{fig:distribution}b) shows the time series of single detector counts in the detector belonging to one ring during a class \emph{III} event. After a sudden start, the rate stabilizes on an elevated level and further increases towards the end of the run. Background measurements with artificially increased pressure ($2\cdot10^{-9}$~mbar) inside the pre-spectrometer - via the injection of high purity argon gas - showed that the rate of class \emph{I} to \emph{III} events is independent of the pressure (see figure \ref{fig:rate5to55h}). However, the duration of a ring event is significantly reduced with increasing pressure, while the total number of counts remains virtually the same and thus, the count rate increases. For another measurement, the magnetic field was lowered to 2.5~\% of the maximum field. In this configuration, no intervals of elevated count rate or ring structures were observed and an average background count rate of (11.8~$\pm$~0.2) $\cdot$ 10$^{-3}$~cps was measured.

\begin{figure}[h]
 \centering
 \includegraphics[width=150mm]{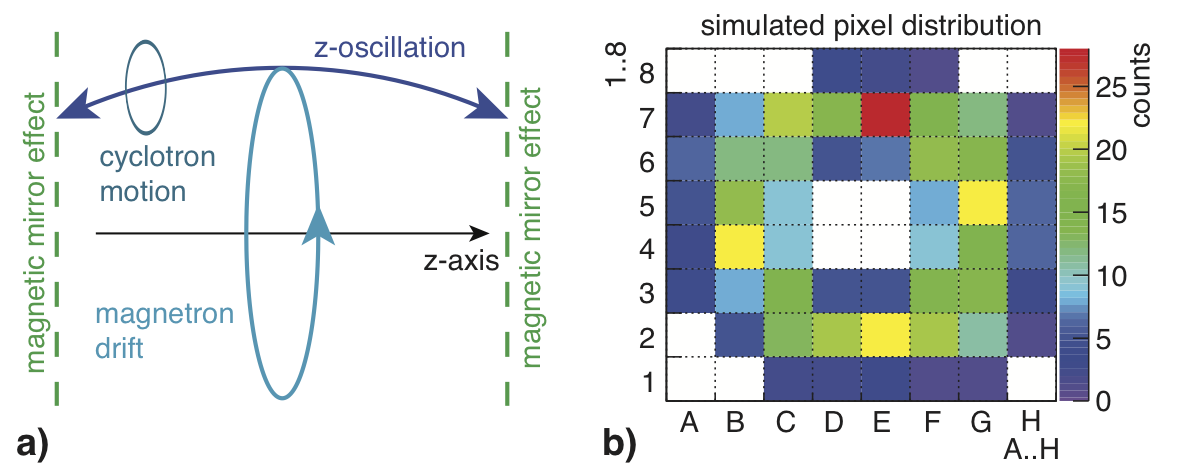}
 \caption{\textbf{a)} Contributions to the motion of a magnetically trapped electron. \textbf{b)} Simulated pixel distribution at the pre-spectrometer detector of secondary electrons produced by a magnetically trapped electron of 178~keV starting energy.}
 \label{fig:simulation}
\end{figure} 

From the measurements described above, it can be concluded that the observed events are due to ionization of residual gas molecules by magnetically trapped high energetic primary electrons. This effect has been observed first in the Troitsk neutrino mass experiment \cite{tro95}. There it was due to tritium $\beta$-decay within the MAC-E filter. This source of background has also been considered in the KATRIN design report \cite{kat04}, yielding an upper limit for the tolerable tritium density in the spectrometer. The effect was further investigated in the MAC-E filter at the Mainz neutrino mass experiment for the case of 17.8~keV conversion electrons from $^{83m}$Kr, decaying inside the spectrometer \cite{bea02}.

The motion of a magnetically trapped electron can be split into a cyclotron motion, an oscillation in $z$-direction caused by magnetic mirroring, and a magnetron drift (see figure \ref{fig:simulation}a)), which is a circular motion - with the spectrometer axis as centre - caused by the inhomogeneous axially symmetric magnetic field ($\vec{B}\times\nabla\vec{B}$) and by crossed electric and magnetic fields ($\vec{E}\times\vec{B}$). Via subsequent ionizations of residual gas molecules, the primary electron creates low energetic secondary electrons that are accelerated by the spectrometer potential and guided via magnetic fields to the detector, where they contribute to the background. Figure \ref{fig:simulation}b) shows the result of a microscopic tracking simulation of a primary electron with an initial energy of 178~keV in the electric and magnetic field configuration of the pre-spectrometer, including scattering processes on hydrogen molecules. The simulated ring structure at the detector is very similar to the ones observed in measurements (see figure \ref{fig:rings}) and can be explained by the magnetron motion of the primary electron. The fuzziness of the ring structure is due to the cyclotron motion of the primary electron and to the change of storage volume of the primary electron due to scattering processes. The simulation also reproduces the time structure of the ring events (see figure \ref{fig:distribution}b)). The increase of rate towards the end of the event can be explained by an increase in the cross section for the ionization of hydrogen molecules by the primary electron as its energy decreases over time. The measurements at higher pressure are also in good agreement with the simulations showing that the background events observed at the pre-spectrometer are most likely due to the storage of high energetic electrons with energies of the order of 1 to 100~keV. Details about the simulations can be found in Reference \cite{mer12}.

\section{Radon background model}
The question was where the high energetic, magnetically trapped electrons are coming from. It seemed unlikely that the electrons originate from the electrode surfaces because magnetic shielding should reflect them back to the electrodes. If this background originated from the electrodes it would also be expected that there are much more events at larger radii. However, the ring events seem to be distributed homogeneously. High energetic cosmic particles and photons can penetrate into the flux tube and could create high energetic electrons via interaction with residual gas molecules. However, those processes should depend on the pressure inside the pre-spectrometer, but increasing the pressure by a factor of 20 did not change the rate of class \emph{I} to \emph{III} events (see figure \ref{fig:rate5to55h}). Only one possibility remained, namely that high energetic electrons are created as a result of the decay of a radioactive gas inside the flux tube of the spectrometer. Taking into account the natural abundance of radioactive isotopes, the most likely candidate was radon from the natural decay chains ($^{219}$Rn, $^{220}$Rn and $^{222}$Rn, see table \ref{tab:radon}).

\subsection{Radon decay chains and decay processes}

\begin{table}[h]
	\caption{Radon isotopes from natural decay chains \cite{bro01}, \cite{wu07}, \cite{jai06}.}
	\label{tab:radon}
		\begin{indented}
		\item[]
		\begin{tabular}{@{}llll}
			\br
			decay chain & Actinium ($^{235}$U) & Thorium ($^{232}$Th) & Radium ($^{238}$U)\\
			\mr
			parent isotope & $^{223}$Ra & $^{224}$Ra & $^{226}$Ra\\
			\bf{radon isotope} & $\mathbf{^{219}}$\bf{Rn} & $\mathbf{^{220}}$\bf{Rn} & $\mathbf{^{222}}$\bf{Rn}\\
			lifetime $\tau_{Rn}$& 5.7 s & 80.5 s & 5.5 d\\
			maximum $\alpha$-energy & 6.8~MeV & 6.3~MeV & 5.5~MeV\\
			daughter isotope & $^{215}$Po & $^{216}$Po & $^{218}$Po\\
			\br
		\end{tabular}
	\end{indented}
\end{table}

The natural abundance of $^{232}$Th, $^{235}$U and $^{238}$U yields three different radioactive decay chains. Each decay chain undergoes a sequence of nuclear decays until a stable lead (Pb) nucleus is created and each decay chain produces its own, characteristic radon isotope (see table \ref{tab:radon}). The decay of radon releases different particles with differing energies, such as electrons, $\alpha$-particle and photons. If a radon atom decays in the vacuum of the pre-spectrometer, the following issues need to be considered:
\begin{enumerate}
	\item $\mathbf{\alpha}$\textbf{-particle:} The maximum energy of the emitted $\alpha$-particle is between 5.5 and 6.8~MeV (see table \ref{tab:radon}).\\
	 Due to its large mass (compared to an electron) and momentum the $\alpha$-particle is practically not influenced by electric and magnetic fields within the pre-spectrometer and, thus, flies on a straight track until it hits the vessel's wall or an inner electrode. Secondary electrons are produced as a result of the impact. However, these electrons are of less concern as they are shielded magnetically.
	\item \textbf{Po recoil:} Radon decays almost exclusively into polonium. The recoil energy of Po is in the range between 107 and 122~keV.\\
	The recoil Po atom can remain a maximum of 12~$\mu$s inside the volume of the pre-spectrometer before it hits the vessel wall or an inner electrode. The probability that it decays during this time is less than 0.5~\%.
	\item \textbf{shake-off electrons:} Any nuclear transition that changes the charge $Z$ or charge distribution within the inner shells of the atom is accompanied by a restructuring process of the atomic shells. In this process, electrons can be emitted. In case of Rn decay up to ten or more (low energetic) secondary electrons can be emitted \cite{szu65}.\\
	The low-energetic shake-off electrons contribute to the pre-spectrometer background as class \emph{0} events.
	\item \textbf{Auger and Coster Kronig electrons:} The $\alpha$-particle emitted from the nucleus can interact with electrons of the atomic shell and eject them, thus creating a vacancy. The probability of this process increases with the atomic shell and is  3.1~\% for the M shell in case of $^{210}$Po $\alpha$-decay \cite{fre74}. The vacancy can be filled with electrons from a higher atomic shell number, and the released energy is emitted as X-ray fluorescence, Auger electrons or Coster-Kronig electrons, with typical energies on the order of keV.\\
	These electrons have a chance of 80 to 90~\% \cite{fra10} of being magnetically trapped inside the pre-spectrometer, and are responsible for the observed class \emph{I} events.
	\item \textbf{conversion electrons:} If an $\alpha$-decay goes to an excited state of the daughter nucleus, there are two competing processes of de-excitation: emission of $\gamma$-rays and internal conversion. In case of internal conversion, the energy is transferred to an electron of the inner shells that then leaves the atom with typical energies of the order of 100~keV. The probability for internal conversion is about 3~\% of each  $^{219}$Rn decay, and electrons with 178~keV (1.27~\%) and 254~keV (0.74~\%) \cite{bro01} are most likely.\\
	Because of the much larger magnetic field in the pre-spectrometer (16~mT at the analyzing plane) - compared to previous experiments in Mainz and Troitsk - the probability of a conversion electron to be trapped inside the pre-spectrometer is close to 1 \cite{fra10}. Due to their high energy, conversion electrons have the potential to create thousands of secondary electrons. Conversion electrons are therefore responsible for the observed class \emph{II} and \emph{III} events.
	\item $\mathbf{\gamma}$\textbf{ and X-rays:} Excitations of the nuclei or vacancies in the atomic shell after the decay of radon yield $\gamma$ and X-rays.\\
	 $\gamma$ and X-rays are not influenced by electric or magnetic fields and produce secondary electrons inside the vessel walls or inner electrodes, which are of less concern regarding background due to the magnetic shielding.
\end{enumerate}

\subsection{Determination of radon activity inside the pre-spectrometer}

\begin{figure}[b]
 \centering
 \includegraphics[width=150mm]{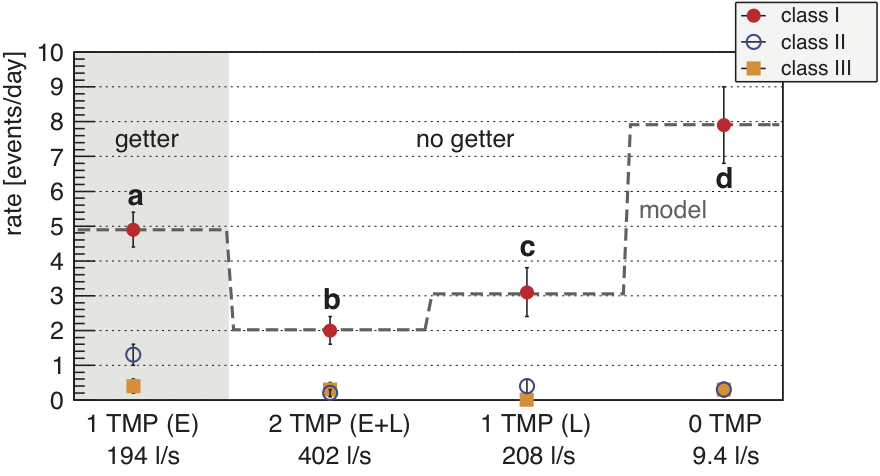}
 \caption{This plot shows the dependence of the rate of class \emph{I} to \emph{III} events on the configuration of the vacuum system (\textbf{a}) with and (\textbf{b, c, d}) without getter pump: \textbf{a} one TMP (Ebara) running, \textbf{b} both TMPs running, \textbf{c} one TMP (Leybold) running, \textbf{d} no TMP running. The dashed line is the result of a fit of the class \emph{I} rates to equation \ref{equ:rnlnequ}.}
 \label{fig:eventsfinalfit}
\end{figure} 

In order to test the hypothesis of a radon induced background in the pre-spectrometer and to determine the radon activity of different radon isotopes, measurements with different configurations of the vacuum system were performed.

If a radon atom is released into the volume of the pre-spectrometer, it can either decay there or be pumped out by the TMPs. Therefore, the probability $P$ of a radon decay inside the pre-spectrometer depends on the lifetime of the radon isotope $\tau_{Rn}$, the pumping speed $S_{Rn}$ and the volume $V$~=~8.5~m$^{3}$ of the vacuum system:

\begin{equation}
	\eqalign{P = \left( 1 + \frac{\tau_{Rn} \cdot S_{Rn}} {V}\right)^{-1}}
	\label{equ:rndecayprob} 
\end{equation}

\begin{table}
	\caption{Radon decay probabilities $P_{i,m}$ [\%] inside the pre-spectrometer for different pumping speeds $S_{Rn}$ using different TMPs (\textbf{E}bara and \textbf{L}eybold).}
	\label{tab:rndecayprob}
		\begin{indented}
		\item[]
		\begin{tabular}{@{}lllll}
			\br
			configuration & $S_{Rn}$ [l/s] & $^{219}$Rn [\%]  & $^{220}$Rn [\%] & $^{222}$Rn [\%]\\
			\mr
			a (E) & 194 & 88.5 & 35.3 & 9.2 $\cdot$ 10$^{-3}$\\
			b (E+L) & 402 & 78.7 & 20.8 & 4.4 $\cdot$ 10$^{-3}$\\
			c (L) & 208 & 87.7 & 33.8 & 8.6 $\cdot$ 10$^{-3}$\\
			d (0) & 9.2 & 99.5 & 96.9 & 2.2\\
			\br
		\end{tabular}
	\end{indented}
\end{table}

The decay probabilities of different radon isotopes for different pumping speeds are shown in table \ref{tab:rndecayprob}. Figure \ref{fig:eventsfinalfit} shows the event rates of class \emph{I} to \emph{III} events for different radon pumping speeds. Class \emph{I} events are an indicator of the total radon activity inside the pre-spectrometer, independent of the isotope. Class \emph{II} and \emph{III} events are an indicator of the $^{219}$Rn activity. In order to obtain quantitative numbers for the radon activities inside the pre-spectrometer, it is assumed that there are three contributions:
\begin{enumerate}
	\item $\mathbf{^{219}}$\textbf{Rn}$\mathbf{_G}$: $^{219}$Rn emanating from the NEG pump
	\item $\mathbf{^{219}}$\textbf{Rn}$\mathbf{_B}$: $^{219}$Rn emanating from the pre-spectrometer or parts attached to it
	\item $\mathbf{^{220}}$\textbf{Rn}$\mathbf{_B}$: $^{220}$Rn emanating from the pre-spectrometer or parts attached to it
\end{enumerate}
It is assumed that a possible $^{222}$Rn contribution can be neglected due to its small decay probability (see table \ref{tab:rndecayprob}) inside the pre-spectrometer. $^{220}$Rn emanation of the NEG material is not taken into account because $\gamma$-ray spectroscopy measurements of the NEG material showed that the activity of the $^{220}$Rn decay chain is two orders of magnitude smaller than the $^{219}$Rn decay chain activity (4.4~Bq/kg). The observed class \emph{I} count rates can be parameterized as a set of linear equations of the form:
\begin{equation}
	\eqalign{r_{I,m} = \sum_{i=1}^3 Rn_i \cdot P_{i,m} \cdot \epsilon_i},
	\label{equ:rnlnequ} 
\end{equation}
where $r_{I,m}$ is the rate of class \emph{I} events for measurement $m$ (\textbf{a} to \textbf{d}), $Rn_i$ is the radon activity of one of the three contributions (see above), $P_{i,m}$ is the decay probability inside the pre-spectrometer (see table \ref{tab:rndecayprob}) and $\epsilon_i$ is the probability that an Auger or Koster Cronig electron is released in the radon decay. The radon activities (see table \ref{tab:rnactivity}) were obtained by applying a fit of equation \ref{equ:rnlnequ} to the data shown in figure \ref{fig:eventsfinalfit}. The good agreement between fit and measured data is a first successful test of the radon hypothesis because equation \ref{equ:rnlnequ} uses the specific lifetimes of the radon isotopes.

\begin{table}
	\caption{Radon activity inside the pre-spectrometer. The measured activity is the observed activity in the detectors field of view which covers about 7~\% of the pre-spectrometer volume. To obtain the total activity, the measured activity was scaled to the pre-spectrometer volume, assuming that the radon decays are distributed homogeneously. The background rate is the average background rate for 100~\% of the KATRIN flux tube and one TMP running.}
	\label{tab:rnactivity}
		\begin{indented}
		\item[]
		\begin{tabular}{@{}llll}
			\br
			contribution & measured activity [mBq] & total activity [mBq] & background [10$^{-3}$cps]\\
			\mr
			$^{219}$Rn$_G$ & 0.55 $\pm$ 0.13 & 7.5 $\pm$ 1.8 & 19 $\pm$ 4 \\
			$^{219}$Rn$_B$ & 0.2 $\pm$ 0.15 & 2.4 $\pm$ 2.0 & 6 $\pm$ 4 \\
			$^{220}$Rn$_B$ & 2.4 $\pm$ 0.7 & 33 $\pm$ 9 & 2.1 $\pm$ 0.4 \\
			\br
		\end{tabular}
	\end{indented}
\end{table}

Based on the radon activities, the rate of class \emph{0} events (single detector counts caused by each radon decay) was estimated and compared with the background rate between the intervals of elevated rate (class \emph{I} to \emph{III} events).  It was found that the prediction is always lower then the observed rate, which is another successful test of the radon hypothesis.

\section{Discussion}

Radon (Rn) atoms, which emanate from materials inside the vacuum region of the KATRIN spectrometers, are able to penetrate deep into the magnetic flux tube so that the $\alpha$-decay of Rn contributes to the background. Of particular importance are electrons emitted in processes accompanying the Rn $\alpha$-decay such as shake-off, internal conversion and Auger electrons. While low-energy electrons directly contribute to the background in the signal region, high-energy electrons can be stored magnetically inside the volume of the spectrometer. Depending on their initial energy, they are able to create thousands of secondary electrons via subsequent ionization processes of residual gas molecules, thus creating a burst of background counts at the detector over a certain period of time. This process depends on the residual gas pressure and on the energy of the stored electron.

For the pre-spectrometer test setup, an average Rn induced background rate of (27~$\pm$~6)~$\cdot$~10$^{-3}$~cps was determined. The emanation of $^{219}$Rn from the getter material was determined to be 7.5~$\pm$~1.8~mBq, thus being responsible for a large fraction ((19~$\pm$~4)~$\cdot$~10$^{-3}$~cps) of the average background rate. The total activity of the $^{219}$Rn decay chain in the NEG material of the installed getter pump was measured in an independent $\gamma$-ray spectroscopy measurement to be about 8~Bq. Hence, the radon emanation efficiency of the NEG material is on the order of 10$^{-3}$.

After removing the NEG pump at the pre-spectrometer, the average background count rate decreased by (19~$\pm$~4) $\cdot$ 10$^{-3}$~cps to (8~$\pm$~4) $\cdot$ 10$^{-3}$~cps. However, sources for $^{219}$Rn (2.4~$\pm$~2.0~mBq) and $^{220}$Rn (33~$\pm$~9~mBq) remained within the test setup. Removing different devices (vacuum gauges, RGAs, glass windows and a thermocouple) from the pre-spectrometer vacuum system reduced the $^{220}$Rn activity by almost an order of magnitude \cite{goe11}.

Based on the results from the KATRIN pre-spectrometer, estimates can be made for the Rn induced background at the KATRIN main spectrometer. Since the main spectrometer will be equipped with 3~km of getter strips, an average background rate - induced of $^{219}$Rn emanating from the getter material - of about 300~$\cdot$~10$^{-3}$~cps \cite{fra10} is expected. This is a factor of 30 larger than the required value of 10~$\cdot$~10$^{-3}$~cps and would reduce the sensitivity of KATRIN on $m_{\overline{\nu}_e}$. In order to reduce the Rn induced background, the installation of LN2 cooled baffles is proposed. The baffles could prevent $^{219}$Rn (emanating from the NEG pumps) from entering the sensitive volume (flux tube) of the main spectrometer and at the same time provide a sufficiently effective pumping speed of the NEG pumps for hydrogen and tritium. Additionally, the baffles could reduce the background rate induced from radon emanating from the main spectrometer vessel itself, or from devices attached to it. In order to determine experimentally the radon suppression efficiency of a LN2 cooled baffle and to test the technical feasibility, a baffle was installed at the pre-spectrometer test setup. With this setup it was shown that the baffle is able to prevent radon emanating from the NEG pump from entering the spectrometer volume. \cite{goe11}.

\end{document}